\documentclass[preprint,superscriptaddress]{revtex4}
\usepackage{graphicx}

\begin{document}

\title{High-frequency suppression of inductive coupling between flux qubit and transmission line resonator}

\author{Sahel Ashhab}
\affiliation{Advanced ICT Institute, National Institute of Information and Communications Technology, 4-2-1, Nukuikitamachi, Koganei, Tokyo 184-8795, Japan}

\author{Ziqiao Ao}
\affiliation{Advanced ICT Institute, National Institute of Information and Communications Technology, 4-2-1, Nukuikitamachi, Koganei, Tokyo 184-8795, Japan}
\affiliation{Department of Applied Physics, Waseda University, Okubo 3-4-1, Shinjuku-ku, Tokyo 169-8555, Japan}
\affiliation{Department of Advanced Science and Engineering, Waseda University, 3-4-1 Okubo, Shinjuku-ku, Tokyo 169-8555, Japan}

\author{Fumiki Yoshihara}
\affiliation{Advanced ICT Institute, National Institute of Information and Communications Technology, 4-2-1, Nukuikitamachi, Koganei, Tokyo 184-8795, Japan}
\affiliation{Department of Physics, Tokyo University of Science, 1-3 Kagurazaka, Shinjuku-ku, Tokyo 162-8601, Japan}

\author{Adrian Lupascu}
\affiliation{Institute for Quantum Computing, University of Waterloo, Waterloo, ON, Canada N2L 3G1}
\affiliation{Department of Physics and Astronomy, University of Waterloo, Waterloo, ON, Canada N2L 3G1}
\affiliation{Waterloo Institute for Nanotechnology, University of Waterloo, Waterloo, ON, Canada N2L 3G1}

\author{Kouichi Semba}
\affiliation{Advanced ICT Institute, National Institute of Information and Communications Technology, 4-2-1, Nukuikitamachi, Koganei, Tokyo 184-8795, Japan}
\affiliation{Institute for Photon Science and Technology, The University of Tokyo, 7-3-1 Hongo, Bunkyo-ku, Tokyo 113-0033, Japan}


\begin{abstract}
We perform theoretical calculations to investigate the naturally occurring high-frequency cutoff in a circuit comprising a flux qubit coupled inductively to a transmission line resonator (TLR). Specifically, a decoupling occurs between the qubit and the high-frequency modes. The coupling strength between the qubit and resonator modes increases with mode frequency $\omega$ as $\sqrt{\omega}$ at low frequencies and decreases as $1/\sqrt{\omega}$ at high frequencies. This result is similar to those of past studies that considered somewhat similar circuit designs. By avoiding the approximation of ignoring the qubit-TLR coupling in certain steps in the analysis, we obtain effects not captured in previous studies. In particular, we obtain a resonance effect that shifts the TLR mode frequencies close to qubit oscillation frequencies. We derive expressions for the TLR mode frequencies, qubit-TLR coupling strengths and qubit Lamb shift. We identify features in the spectrum of the system that can be used in future experiments to test and validate the theoretical model.
\end{abstract}

\maketitle

\section{Introduction}
\label{Sec:Introduction}

The fields of cavity quantum electrodynamics (cavity-QED) \cite{Andreev,Walls,Scully,Gerry,Glushkov} and circuit quantum electrodynamics (circuit-QED) \cite{Bishop,Blais,Fabre} have proved to be ubiquitous and important in the development of physics for the past few decades. In particular, advances in superconducting circuit technology have allowed the development of qubit-oscillator systems in the ultrastrong- and deep-strong-coupling regimes \cite{Niemczyk,FornDiaz2010,FornDiaz2017,Yoshihara2017NP,Yoshihara2017PRA,Rossatto,FornDiaz2019}. Another related development is the design of strong coupling between superconducting qubits and multimode resonators \cite{Sundaresan,Bosman,Chakram,Ao,Ann}.

As we shall discuss in more detail below, a complication arises in the theoretical treatment of the high-frequency modes in a multimode resonator. In the perturbative approach in which the qubit-resonator coupling is treated as a perturbation to the bare qubit and resonator Hamiltonians, high-frequency divergences arise. In particular, the coupling strength between the qubit and individual modes grows indefinitely with mode frequency. Because the Lamb shift depends on the ratio between the coupling strength and the mode frequency, and the coupling strength increases only as the square-root of the mode frequency, the Lamb shift caused by individual modes decreases with increasing mode frequency. However, the decrease is slow, such that the total Lamb shift diverges. This complication was avoided in past theoretical studies by imposing a frequency cutoff above which resonator modes are ignored. It was recently shown, however, that no ad-hoc cutoff is needed. A natural decoupling  occurs between the qubit and high-frequency modes, which eliminates the divergences mentioned above \cite{Gely,Malekakhlagh2017,ParraRodriguez}. It is worth noting that there have been other studies on the high-frequency cutoff in superconducting circuits \cite{Malekakhlagh2016,McKay,Shi,Roth,Hassler} and atom-cavity systems \cite{DeLiberato,DeBernardis,Ashida}.

In this work, we present a theoretical treatment of a circuit-QED system that comprises a flux qubit coupled inductively to a quarter-wavelength transmission line resonator (TLR). The circuit was investigated experimentally in Ref.~\cite{Ao}. Our calculations are complementary to those of Refs.~\cite{Gely,Malekakhlagh2017}, which focused on charge and transmon qubits coupled capacitively to half-wavelength TLRs, and that of Ref.~\cite{ParraRodriguez}, which provided a general framework for treating circuit-QED systems that contain multimode resonators. Similarly to past studies, we find a natural decoupling at high frequencies. By avoiding certain approximations, we obtain effects not captured in previous studies, including a resonance effect between qubit and TLR frequencies. As we go along in our analysis, we derive a variety of formulae for the TLR normal modes, the coupling strength and the Lamb shift.

\section{Circuit, Lagrangian and Hamiltonian}
\label{Sec:Setup}

\begin{figure}[h]
\includegraphics[width=14.0cm]{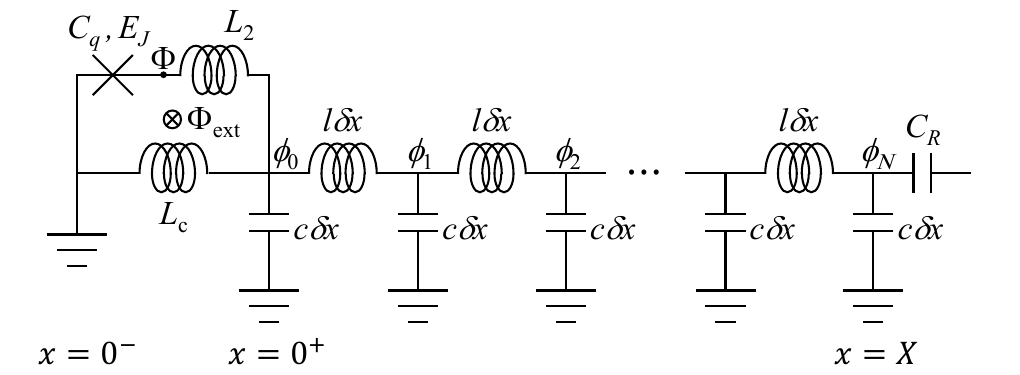}
\caption{Schematic diagram of the qubit-TLR system, with the TLR modeled as a series of LC elements. After setting up the equations with a finite length $\delta x$, we will take the limit $\delta x \to 0$. The TLR is a quarter-wavelength resonator, terminated to ground at one end and with a (near-zero-capacitance) capacitor at the other end. The qubit is modeled as a loop with a single Josephson junction, in addition to linear inductances. The loop is threaded by an externally applied flux $\Phi_{\rm ext}$. The qubit-TLR coupling arises from the shared inductance $L_c$. The $x$ values at the bottom show the correspondence with the original circuit and mathematical model.}
\label{Fig:CircuitDiagram}
\end{figure}

\begin{figure}[h]
\includegraphics[width=12.0cm]{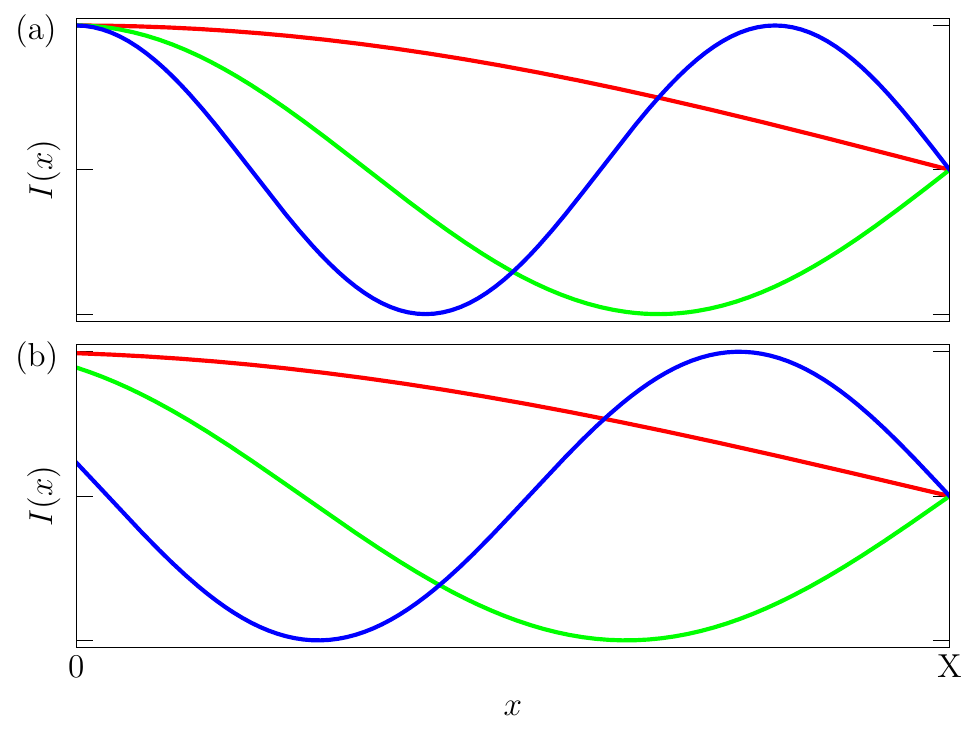}
\caption{Electric current profiles $I(x)$ of the lowest three normal modes of the quarter-wavelength TLR in the absence of coupling to the qubit (a) and in the case of coupling to the qubit (b). Note that, apart from the profile of the current as a function of $x$ for each mode, the maximum amplitudes in the figure do not have any physical significance; for example, they should not be compared to each other. The point $x=0$ is where the TLR makes contact with the ground plane. In the absence of the qubit (Panel a), there is no charge accumulation, and all the modes have horizontal current profiles at $x=0$. The presence of the qubit at $x=0$ (Panel b) interrupts the current flow at the location of the qubit, which suppresses the current amplitudes at that location. As with a low-pass filter, the suppression effect becomes increasingly strong with increasing mode frequency, with $I(0)\to 0$ in the limit of infinitely high mode frequency. The amplitude suppression is exaggerated in this figure for clarity.}
\label{Fig:ModeDiagram}
\end{figure}

We consider the circuit shown in Fig.~\ref{Fig:CircuitDiagram}. Compared to the experimental circuit used in Ref.~\cite{Ao}, we make the standard simplification where we keep only one Josephson junction in the theoretical treatment. This one junction is sufficient to provide the anharmonicity needed to create a qubit. Two other (large) junctions in the qubit loop are replaced by a single inductance. The coupling junction that is shared by the qubit loop and the resonator is also replaced by a coupling inductance. As shown in Ref.~\cite{Yoshihara2022}, these approximations can provide a rather accurate theoretical treatment of circuit-QED systems even in the deep-strong-coupling regime. To maximize the coupling between the qubit and TLR, the TLR is shorted to the ground at the end where the qubit is located. In this case, all TLR normal modes have maximum current amplitudes at the location of the qubit, as illustrated by the electric current profiles in Fig.~\ref{Fig:ModeDiagram}(a). The other end of the TLR is terminated by a capacitor. We replace the spatially extended TLR by a long series of small inductances and capacitors, which is a standard approach to model such resonators. Each inductance-capacitor pair corresponds to a segment of length $\delta x$ of the TLR. At some point in our derivations below, we will take the limit in which the inductance-capacitance series is replaced by an infinite series of infinitesimally small elements, corresponding to the infinitesimal length $dx$.

We follow the standard technique for calculating the quantum mechanical Hamiltonian for superconducting circuits \cite{Devoret,Vool}. From the circuit diagram shown in Fig.~\ref{Fig:CircuitDiagram}, we can construct the Lagrangian:
\begin{eqnarray}
\mathcal{L} & = & \frac{1}{2} C_q \dot{\Phi}^2 - U_q(\Phi,\Phi_{\rm ext}) + \sum_{j=0}^{N} \frac{1}{2} c \delta x \dot{\phi}_j^2 - \sum_{j=0}^{N-1} \frac{1}{2l \delta x} \left( \phi_j - \phi_{j+1} \right)^2 \nonumber \\ & & \hspace{6cm} - \frac{1}{2L_c} \phi_0^2 - \frac{1}{2L_2} \left( \Phi - \phi_0 \right)^2 + \frac{1}{2} C_R \dot{\phi}_N^2.
\end{eqnarray}
The flux (or phase) variable $\Phi$ represents the qubit variable that introduces the nonlinearity in the circuit via the effective Josephson potential $U_{\rm{q}}(\Phi,\Phi_{\rm ext})= - E_J \cos \left\{ {2 \pi (\Phi- \Phi_{\rm ext})/\Phi_0 } \right\}$. The coefficient $E_J=I_c \Phi_0 / (2\pi)$ is the Josephson energy of the junction, $I_c$ is the critical current, and $\Phi_0=h/(2e)$ is the superconducting flux quantum. The variables $\phi_j$ are the phase variables at the nodes $j$ ($j=0,1,2,...,N$) along the TLR. The capacitance $C_q$ is the junction capacitance. The parameters $c$ and $l$ are, respectively, the capacitance and inductance per unit length of the TLR, such that the capacitances and inductances of the elements shown in Fig.~\ref{Fig:CircuitDiagram} are given by $c\delta x$ and $l\delta x$.

Using the Legendre transformation,
\begin{eqnarray}
Q & = & \frac{\partial \mathcal{L}}{\partial \dot{\Phi}} = C_q \dot{\Phi} \\
\delta q_j & = & \frac{\partial \mathcal{L}}{\partial \dot{\phi}_j} = c \delta x \dot{\phi}_j, \ 0 \leq j \leq N-1 \\
\delta q_N & = & \frac{\partial \mathcal{L}}{\partial \dot{\phi}_N} = (C_R+c \delta x) \dot{\phi}_N,
\end{eqnarray}
we obtain the Hamiltonian:
\begin{eqnarray}
\mathcal{H} & = & \frac{1}{2 C_q} Q^2 + U_q(\Phi,\Phi_{\rm ext}) + \sum_{j=0}^{N-1} \left( \frac{1}{2c \delta x} \delta q_j^2 + \frac{1}{2l \delta x} \left( \phi_j - \phi_{j+1} \right)^2 \right) \nonumber \\ & & \hspace{5cm} + \frac{1}{2L_c} \phi_0^2 + \frac{1}{2L_2} \left( \Phi - \phi_0 \right)^2 + \frac{1}{2 (C_R+c\delta x)} \delta q_N^2.
\label{Eq:HamiltonianCV}
\end{eqnarray}
As is common in the theoretical treatment of superconducting circuits, the main changes that occur when going from the Lagrangian to the Hamiltonian are: (1) the time derivatives of the phase variables are replaced by charge variables, (2) capacitances move from numerators to denominators, and (3) the signs of the potential energy terms are reversed. The smallness of $\delta x$ adds the following feature in our case: the new variables $\delta q_j$ are proportional to $\delta x$. Hence, the term $\delta q_j^2/(2c\delta x)$ is also proportional to $\delta x$. Similarly, since the difference $\phi_j - \phi_{j+1}$ is equal to the spatial derivative of the phase variable multiplied by $\delta x$, the term $\left( \phi_j - \phi_{j+1} \right)^2 / (2l \delta x)$ is proportional to $\delta x$. These properties are to be expected and allow a smooth transition from this discrete description to the continuous description that we shall introduce shortly.

It is worth pointing out another interesting property of the Hamiltonian: although $L_c$ is the mutual inductance between the qubit and the TLR, it is $L_2$ that appears in the Lagrangian and Hamiltonian terms that combine the qubit and TLR variables, i.e.~the variables $\Phi$ and $\phi_0$. Furthermore, $L_2$ appears in the denominator, whereas we expect the mutual inductance to appear in the numerator in the effective coupling term, which should be equal to the mutual inductance times the product of the qubit and TLR currents. This situation might seem paradoxical. However, it is simply a matter of deceptive appearance in the intermediate steps of the derivation. For example, a somewhat similar situation was encountered in Ref.~\cite{Yoshihara2022}. In that case a Y-$\Delta$ transformation produces a coupling term that has the coupling inductance $L_c$ in the numerator, as intuitively expected. It is less obvious how a similar transformation could give rise to a logical-looking coupling term in the present case. We will show in Appendix B how a careful analysis of the Hamiltonian produces logical results. For the time being, we just point out that if we set $L_c=0$, we find that we must have $\phi_0=0$ to avoid a divergence in the $\phi_0^2/(2L_c)$ term, which in turn causes the term $\Phi\phi_0/L_2$ to vanish. Hence we indirectly obtain the expected result that the value $L_c=0$ corresponds to decoupled subsystems.

The question of the qubit-TLR coupling term form also leads to the question of gauge choice. The Hamiltonian in Eq.~(\ref{Eq:HamiltonianCV}) is in the flux gauge, which is characterized by a coupling term containing a product of phase variables. The Hamiltonian can be transformed via a gauge transformation to the charge gauge, in which the coupling term contains a product of charge variables. Using the unitary operator
\begin{equation}
\mathcal{U} = \exp \left\{ \frac{1}{i\hbar} \nu \Phi \sum_{j=0}^{N} \delta q_j \right\},
\end{equation}
where $\nu$ is a constant, we obtain the transformed Hamiltonian
\begin{eqnarray}
\mathcal{H}' & = & \mathcal{U}^{\dagger} \mathcal{H} \mathcal{U}
\nonumber \\
& = & \frac{1}{2 C_q} \left( Q - \nu \sum_{j=0}^{N} \delta q_j \right)^2 + U_q(\Phi,\Phi_{\rm ext}) + \sum_{j=0}^{N-1} \left( \frac{1}{2c \delta x} \delta q_j^2 + \frac{1}{2l \delta x} \left( \phi_j - \phi_{j+1} \right)^2 \right) \nonumber \\ & & \hspace{3.5cm} + \frac{1}{2L_c} \left( \phi_0 + \nu \Phi \right)^2 + \frac{1}{2L_2} \left( \Phi - \phi_0 - \nu \Phi \right)^2 + \frac{1}{2 (C_R+c\delta x)} \delta q_N^2.
\label{Eq:HamiltonianTransformed}
\end{eqnarray}
The charge gauge is obtained by setting $\nu = L_c/(L_c+L_2)$, which eliminates the terms that contain the product $\Phi\phi_0$. Instead, the qubit-TLR coupling is now described by the product $Q \times \sum_{j=0}^{N} \delta q_j$. Considering that the coupling term now involves long-range interactions between the qubit's charge variable $Q$ and the charge densities at all points in the TLR, it is simpler to work in the flux gauge. This result is consistent with past results in the literature \cite{DeBernardis,Yoshihara2022}, which found that it is simpler to use the flux gauge when dealing with circuits that comprise flux qubits coupled to resonators.

\section{Equations of motion and boundary conditions}
\label{Sec:Equations}

From the Hamiltonian, we can derive the equations of motion for the dynamical variables straightforwardly:
\begin{eqnarray}
\dot{Q} & = & - \frac{\partial \mathcal{H}}{\partial \Phi} = - \frac{dU_q(\Phi,\Phi_{\rm ext})}{d\Phi} - \frac{\Phi - \phi_0}{L_2} \\
\dot{\Phi} & = & \frac{\partial \mathcal{H}}{\partial Q} = \frac{Q}{C_q} \\
\dot{\delta q}_j & = & - \frac{\partial \mathcal{H}}{\partial \phi_j} = \left\{ \begin{array}{lll}
\frac{\phi_1 - \phi_0}{l\delta x} - \frac{\phi_0}{L_c} + \frac{\Phi - \phi_0}{L_2}, & & j = 0, \\
\frac{\phi_{j+1} - 2\phi_j + \phi_{j-1}}{l\delta x}, & & 1 \leq j \leq N-1, \\
\frac{-\phi_N + \phi_{N-1}}{l\delta x}, & & j = N,
\end{array} \right. \\
\dot{\phi}_j & = & \frac{\partial \mathcal{H}}{\partial \delta q_j} = \left\{ \begin{array}{lll}
\frac{\delta q_j}{c\delta x}, & & 0 \leq j \leq N-1, \\
\frac{\delta q_N}{C_R+c\delta x}, & & j = N.
\end{array} \right.
\end{eqnarray}

It is convenient for the analysis below to turn the first-order equations of motion into second-order equations of motion for the phase variables
\begin{eqnarray}
\ddot{\Phi} & = & - \frac{1}{C_q} \frac{dU_q(\Phi,\Phi_{\rm ext})}{d\Phi} - \frac{\Phi - \phi_0}{C_q L_2}
\label{Eq:2ndOEOMDiscretePhi}
\\
\ddot{\phi}_j & = & \left\{ \begin{array}{lll}
\frac{\phi_1 - \phi_0}{cl\delta x^2} - \frac{\phi_0}{c\delta x L_c} + \frac{\Phi - \phi_0}{c\delta x L_2}, & & j = 0, \\
\frac{\phi_{j+1} - 2\phi_j + \phi_{j-1}}{cl\delta x^2}, & & 1 \leq j \leq N-1, \\
\frac{- \phi_N + \phi_{N-1}}{l\delta x(C_R+c\delta x)}, & & j = N.
\end{array} \right.
\label{Eq:2ndOEOMDiscretephi}
\end{eqnarray}

We now take the limit in which $\delta x$ becomes the infinitesimal $dx$, and we obtain the continuous version of Eqs.~(\ref{Eq:2ndOEOMDiscretePhi},\ref{Eq:2ndOEOMDiscretephi}):
\begin{eqnarray}
\ddot{\Phi} & = & - \frac{1}{C_q} \frac{dU_q(\Phi,\Phi_{\rm ext})}{d\Phi} - \frac{\Phi - \phi(x=0^+,t)}{C_q L_2}
\label{Eq:2ndOEOMContinuousPhi}
\\
\frac{\partial^2 \phi}{\partial t^2} & = & \left\{ \begin{array}{lll}
\left( \frac{1}{cl} \frac{\partial \phi}{\partial x} - \frac{\phi}{c L_c} + \frac{\Phi - \phi}{c L_2} \right) \delta(x), & & x = 0, \\ 
\frac{1}{cl} \frac{\partial^2 \phi}{\partial x^2}, & & x > 0. \end{array} \right.
\label{Eq:2ndOEOMContinuousphi}
\end{eqnarray}
In Eq.~(\ref{Eq:2ndOEOMContinuousphi}), $\delta (x)$ is the Dirac delta function, which is the natural limit for the factor $1/\delta x$, since the Dirac delta function can be thought of as an extremely narrow step-function peak whose height is the inverse of its width. From a different point of view, the role of the $j=0$ or $x=0^+$ point is to set the effective boundary condition for the phase variable in the TLR just to the right of the qubit in Fig.~\ref{Fig:CircuitDiagram}. The Dirac delta function naturally plays this role, as we shall see shortly.

Since the TLR is connected to the ground at $x=0$ (to the left of the qubit), the boundary condition there is
\begin{eqnarray}
\phi(x,t)\Big|_{x=0^-} & = & 0.
\end{eqnarray}
If the TLR is well isolated from the environment on the right-hand side of Fig.~\ref{Fig:CircuitDiagram}, the capacitance $C_R$ will be small. For purposes of this argument, we can take the limit $C_R\to 0$. Considering that $(- \phi_N + \phi_{N-1})/\delta x=-(\partial \phi/\partial x)|_{x=X}$, where $X$ is the length of the TLR, we find that to avoid a divergence in the last line of Eq.~(\ref{Eq:2ndOEOMDiscretephi}) the boundary condition at $x=X$ must be
\begin{eqnarray}
\frac{\partial \phi}{\partial x} \Bigg|_{x=X} & = & 0.
\end{eqnarray}
In physical terms, this boundary condition means that the current at $x=X$ is equal to zero, which is needed to prevent the accumulation of an infinite charge density at that point.

By integrating Eq.~(\ref{Eq:2ndOEOMContinuousphi}) from $x=0^-$ to $x=0^+$, we rewrite the equations of motion in the simpler form:
\begin{eqnarray}
\ddot{\Phi} & = & - \frac{1}{C_q} \frac{dU_q(\Phi,\Phi_{\rm ext})}{d\Phi} - \frac{\Phi - \phi(x=0,t)}{C_q L_2}
\label{Eq:PhiPreRedefinition} \\
\frac{\partial^2 \phi}{\partial t^2} & = & \frac{1}{cl} \frac{\partial^2 \phi}{\partial x^2},
\label{Eq:phiPreRedefinition}
\end{eqnarray}
with modified boundary conditions:
\begin{eqnarray}
\left( \frac{1}{cl} \frac{\partial \phi}{\partial x} - \frac{\phi}{c L_c} + \frac{\Phi - \phi}{c L_2} \right) \Bigg|_{x=0} & = & 0 \label{Eq:BoundaryCondition0} \\
\frac{\partial \phi}{\partial x} \Bigg|_{x=X} & = & 0.
\label{Eq:BoundaryConditionXPreRedefinition}
\end{eqnarray}

\section{TLR modes and high-frequency decoupling}
\label{Sec:Solutions}

We now calculate the frequencies and electric current profiles of the normal modes in the TLR. First, we note that the presence of $\Phi$ in the boundary condition in Eq.~(\ref{Eq:BoundaryCondition0}) means that in principle the equations for $\phi$ cannot be solved without knowledge of $\Phi$. We can however, eliminate $\Phi$ from the equations for $\phi$ under the approximation that $\Phi$ exhibits only small dynamical deviations away from its mean value $\overline{\Phi}$, which is a reasonable assumption for the low-energy states of the system. A detailed derivation on this point is given in Appendix A. If we assume that $\Phi$ remains identically equal to $\overline{\Phi}$ at all times, and we define the new variable
\begin{equation}
\tilde{\phi}(x,t) = \phi(x,t) - \frac{L_{c2}}{L_2} \overline{\Phi},
\end{equation}
where $L_{c2}=L_c L_2 / (L_c + L_2)$, we obtain $\Phi$-independent equations for $\tilde{\phi}$. The equations are given by
\begin{eqnarray}
\frac{\partial^2 \tilde{\phi}}{\partial t^2} & = & \frac{1}{cl} \frac{\partial^2 \tilde{\phi}}{\partial x^2}
\label{Eq:phiPostRedefinition} \\
\left( \frac{1}{cl} \frac{\partial \tilde{\phi}}{\partial x} - \frac{\tilde{\phi}}{c L_{c2}} \right) \Bigg|_{x=0} & = & 0
\label{Eq:BoundaryCondition0PostRedefinition} \\
\frac{\partial \tilde{\phi}}{\partial x} \Bigg|_{x=X} & = & 0.
\label{Eq:BoundaryConditionXPostRedefinition}
\end{eqnarray}

We can now determine the normal modes by solving this one-dimensional wave equation. We do so by substituting sinusoidally oscillating solutions with temporal frequency $\omega$:
\begin{eqnarray}
\tilde{\phi}(x,t) = e^{-i\omega t} u(x),
\end{eqnarray}
which gives
\begin{eqnarray}
- \omega^2 u(x) & = & \frac{1}{cl} \frac{d^2 u}{d x^2}
\label{Eq:WaveEquationU} \\
\left( \frac{1}{cl} \frac{d u}{d x} - \frac{u}{c L_{c2}} \right) \Bigg|_{x=0} & = & 0 \\
\frac{d u}{d x} \Bigg|_{x=X} & = & 0.
\end{eqnarray}

\begin{figure}[h]
\includegraphics[width=12.0cm]{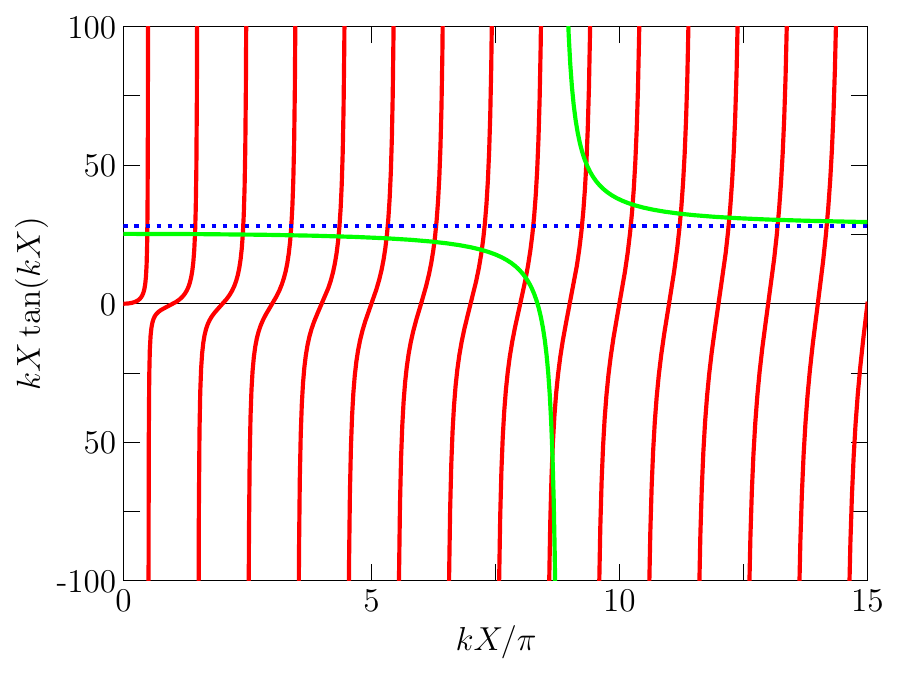}
\caption{The function $kX \tan(kX)$, shown in red, as a function of $kX/\pi$. The horizontal blue dotted line represents the constant on the right-hand side of Eq.~(\ref{Eq:NormalModeEquation_k}). Here it is set to $Xl/L_{c2}=28$, which corresponds to the parameters of Ref.~\cite{Ao}. The points of intersection between the red and blue lines correspond to solutions of Eq.~(\ref{Eq:NormalModeEquation_k}). For the low-frequency modes, i.e.~at small values of $kX$, these intersection points are close to half-integer values of $kX/\pi$. For the high-frequency modes, as we go to increasingly large values of $kX$, the intersection points become increasingly close to integer values of $kX/\pi$. The green line replaces the blue line in the more precise calculation presented in Appendix A and Eq.~(\ref{Eq:NormalModeEquationIncludingPhiTerm}), with the parameters $\omega_q \sqrt{cl}X/\pi =8.8$ and $\mu=0.1$. The intersection points between the red and green lines give the solutions to Eq.~(\ref{Eq:NormalModeEquationIncludingPhiTerm}). When compared with the blue line, the green line gives small corrections to most of the intersection points and a resonance feature around $k=\omega_q \sqrt{cl}$. The resonance feature includes one additional intersection point that corresponds to a qubit oscillation mode, as well as significant shifts in the TLR mode frequencies.}
\label{Fig:ModeFrequencyTranscendentalEquation}
\end{figure}

The solution of the wave equation in Eq.~(\ref{Eq:WaveEquationU}) can be expressed as
\begin{eqnarray}
u(x) = u_c \cos (k[x-x_0]),
\end{eqnarray}
where $k=\sqrt{\omega^2 cl}$. The parameters $k$ and $x_0$ are determined by the boundary conditions. It is worth noting here that, since the differential equation and boundary conditions are linear, they do not impose any conditions on $u_c$. As we will see below, since $u_c$ is the amplitude of current oscillations, it will be governed by energy quantization, especially when analyzing the ground state of the system. Considering the boundary condition at $x=X$, we can write $u(x)$ as
\begin{eqnarray}
u(x) = u_c \cos (k[x-X]).
\end{eqnarray}
The boundary condition at $x=0$ now gives
\begin{eqnarray}
\frac{k}{cl} \sin(kX) - \frac{1}{c L_{c2}} \cos(kX) = 0,
\end{eqnarray}
and therefore
\begin{equation}
k \tan(kX) = \frac{l}{L_{c2}}.
\label{Eq:NormalModeEquation_k}
\end{equation}
This transcendental equation has an infinite number of solutions, i.e.~there are an infinite number of $k$ values that satisfy the equation, as illustrated in Fig.~\ref{Fig:ModeFrequencyTranscendentalEquation}. These solutions characterize the TLR modes. As we will show with a detailed analysis below, the mode current profiles are modified from the form illustrated in Fig.~\ref{Fig:ModeDiagram}(a) to that illustrated in Fig.~\ref{Fig:ModeDiagram}(b).

As explained in Appendix A, including the first-order corrections caused by the deviations of $\Phi$ from $\overline{\Phi}$ leads to replacing Eq.~(\ref{Eq:NormalModeEquation_k}) by
\begin{equation}
k \tan(kX) = \frac{l}{L_{c2}} \times \left( 1 + \frac{\mu}{\left( \frac{k^2}{cl\omega_q^2} - 1 \right)} \right),
\label{Eq:NormalModeEquationIncludingPhiTerm}
\end{equation}
where $\mu=L_c L_q/[(L_c+L_2)(L_q+L_2)]$, and $\omega_q$ is a characteristic qubit oscillation frequency defined in Appendix A. This equation is clearly more cumbersome than Eq.~(\ref{Eq:NormalModeEquation_k}). We shall therefore continue the derivations below using Eq.~(\ref{Eq:NormalModeEquation_k}), and we shall introduce the corrections that arise from using Eq.~(\ref{Eq:NormalModeEquationIncludingPhiTerm}) only when needed.

It is helpful at this point to rewrite Eq.~(\ref{Eq:NormalModeEquation_k}) in the form
\begin{equation}
\omega \tan(kX) = \omega_{\rm cutoff},
\label{Eq:NormalModeEquation_omega_k}
\end{equation}
or alternatively,
\begin{equation}
\omega \tan(\omega \sqrt{cl} X) = \omega_{\rm cutoff},
\end{equation}
where
\begin{equation}
\omega_{\rm cutoff} = \frac{Z_0}{L_{c2}},
\end{equation}
and $Z_0=\sqrt{l/c}$ is the impedance of the TLR, typically set to 50 $\Omega$. This expression for $\omega_{\rm cutoff}$ is the one for an inductance-based low-pass filter. This result is in line with that of Ref.~\cite{Malekakhlagh2017}, where the authors treated capacitive qubit-TLR coupling and found a cutoff frequency that corresponds to a capacitor-based low-pass filter. We emphasize, however, that in spite of the similarity in the underlying physics, the circuit of Ref.~\cite{Malekakhlagh2017} and our circuit are substantially different and cannot be transformed into each other for example via a gauge transformation. Similar expressions for the cutoff frequency also appear in Refs.~\cite{Gely,ParraRodriguez,Shitara}. The physical meaning of $\omega_{\rm cutoff}$ will become clear when we discuss the high-frequency limit below.

\subsection{Low-frequency modes}

For the low-frequency modes with $\omega\ll\omega_{\rm cutoff}$, the solutions of Eq.~(\ref{Eq:NormalModeEquation_omega_k}) must have $\tan(kX)\gg 1$, which implies that $kX$ is slightly smaller than $n\pi+\pi/2$ with $n$ being an integer. We therefore define the small variable $\widetilde{kX} = n\pi + \pi/2 - kX$. By rearranging Eq.~(\ref{Eq:NormalModeEquation_k}) and making use of the first-order approximation $\cot(n\pi+\pi/2-\delta)\approx\delta$ for $\delta\ll 1$, we obtain
\begin{eqnarray}
n \pi + \frac{\pi}{2} - \widetilde{kX} & = & \frac{Xl}{L_{c2}} \cot (n\pi + \frac{\pi}{2} - \widetilde{kX})
\nonumber \\
& \approx & \frac{Xl}{L_{c2}} \widetilde{kX}.
\label{Eq:LowFrequencyNormalModeEquation}
\end{eqnarray}
As a result,
\begin{equation}
\widetilde{kX} \approx \frac{n \pi + \frac{\pi}{2}}{1 + Xl/L_{c2}}.
\end{equation}
Taking into consideration the fact that $Xl/L_{c2}\gg 1$ (which is required for the validity of the condition $\omega\ll\omega_{\rm cutoff}$), we obtain the approximation
\begin{equation}
kX \approx n \pi + \frac{\pi}{2} - \frac{\left( n \pi + \frac{\pi}{2} \right)}{1+Xl/L_{c2}} \approx \left( n \pi + \frac{\pi}{2} \right) \times \left(1 - \frac{L_{c2}}{Xl} \right),
\label{Eq:LowFrequencySolutionToFirstOrder}
\end{equation}
starting at $n=0$ for the fundamental mode. From now on, we shall refer to the value of $k$ that corresponds to a certain value of $n$ as $k_n$, and we shall refer to the corresponding frequency as $\omega_n$.

\subsection{Fundamental mode}

Using various expressions given earlier in this section, we find the (zeroth-order) approximate expression for the fundamental mode frequency
\begin{equation}
\omega_0 = \frac{\pi}{2X\sqrt{cl}} = \frac{\pi Z_0}{2Xl} = \frac{\pi}{2X Z_0 c},
\label{Eq:FundamentalModeFrequency}
\end{equation}
as should be expected for a quarter wavelength resonator. It is worth noting that with this expression for $\omega_0$, we find that $Xl/L_{c2}=\pi\omega_{\rm cutoff}/(2\omega_0)$, and Eq.~(\ref{Eq:NormalModeEquation_k}) can be rewritten as
\begin{equation}
k_n X \tan(k_n X) = \frac{\pi\omega_{\rm cutoff}}{2\omega_0}.
\end{equation}
It is also worth noting that the factor $(1-L_{c2}/Xl)$ in Eq.~(\ref{Eq:LowFrequencySolutionToFirstOrder}) is approximately equal to $Xl/(Xl+L_{c2})$. Using the expression $\omega_0=\pi Z_0/(2Xl)$, the small deviation of the factor $Xl/(Xl+L_{c2})$ from unity can be understood in terms of the inductance $L_{c2}$ increasing the total inductance of the TLR from $Xl$ to $Xl+L_{c2}$.

\subsection{High-frequency modes}

For the high-frequency modes with $\omega\gg\omega_{\rm cutoff}$, the solutions of Eq.~(\ref{Eq:NormalModeEquation_k}) must have $\tan(k_n X)\ll 1$, which means that $k_n X$ must be slightly larger than $n\pi$, with $n$ being an integer. We therefore define $\widetilde{k_n X} = k_n X - n\pi$, which gives
\begin{equation}
(n\pi + \widetilde{k_n X}) \tan(n\pi + \widetilde{k_n X}) = \frac{Xl}{L_{c2}}.
\end{equation}
Using the approximation $\tan(n\pi+\delta)\approx\delta$ for $\delta\ll 1$, we obtain the first-order approximation in $\omega_{\rm cutoff}/\omega$
\begin{equation}
\widetilde{k_n X} = \frac{Xl}{n\pi L_{c2}}.
\end{equation}
This formula in turn gives the approximate expression for $k_n X$:
\begin{equation}
k_n X \approx n \pi + \frac{Xl}{n\pi L_{c2}}.
\end{equation}

Note that in the limit $n\rightarrow\infty$, we obtain $k_n X=n\pi$, and the boundary condition at $x=0$ effectively becomes $du/dx=0$, which is the appropriate boundary condition if the TLR were terminated with capacitors at both ends, i.e.~the connection to the ground is effectively cut and no (high-frequency) current flows at the point where the qubit is located. The low-pass-filter behavior now becomes clear. In fact, this behavior is perhaps more intuitive for the circuit with inductive coupling compared to the case of capacitive coupling studied in Ref.~\cite{Malekakhlagh2017}. The intuitive picture of an inductance is as a circuit element that resists changes in current. The time derivative of an ac current in the inductance is given by the product of the oscillation amplitude and the frequency. Put differently, the impedance of an inductance $L$ at frequency $\omega$ is given by $i\omega L$. Increasing the frequency leads to stronger resistance by the inductance, which can be alleviated by suppressing the oscillation amplitude. In the limit of infinite frequency, the current at the location of the low-pass filter must be suppressed to zero to avoid an infinite resistance from the inductance.

In addition to $\omega_{\rm cutoff}$, we can also define the parameter $n_{\rm cutoff}$ as an estimate for the number of TLR modes whose coupling to the qubit is not significantly suppressed:
\begin{eqnarray}
n_{\rm cutoff} = \frac{\omega_{\rm cutoff}}{\omega_0} = \frac{2Xl}{\pi L_{c2}}.
\end{eqnarray}
The actual number of modes below the cutoff frequency is $n_{\rm cutoff}/2$. The factor of 2 in the denominator arises because the fundamental mode ($n=0$) has $k_0 X=\pi/2$ while the distance between adjacent modes is $(k_{n+1}-k_n) X\approx \pi$. Another way to look at the factor of 2 is to note that the normal mode frequencies of a quarter-wavelength resonator are odd multiples of the fundamental mode frequency, and the even multiples are missing. It should be emphasized, however, that this factor does not have too much significance, because the current suppression with increasing mode frequency is a gradual process and not a sharp cutoff.

\subsection{Coupling strengths between qubit and TLR normal modes}

Now that we have determined the frequencies and current profiles of the normal modes in the TLR, we can calculate their quantum properties and determine how they interact with the qubit. From the form of the Hamiltonian, we can see that the energy is proportional to $u_c^2$:
\begin{equation}
E \approx \frac{Xk_n^2u_c^2}{2l}.
\end{equation}
In the ground state of any of the modes, the energy should be $E=\hbar\omega_n/2$, which is the ground-state energy of a harmonic oscillator. Equating these two formulae for the energy gives the formula for the zero-point (root-mean-square) fluctuations in the mode variable:
\begin{equation}
u_{c, \rm rms} = \sqrt{\frac{\hbar}{Xc\omega_n}}.
\end{equation}
These fluctuations give the zero-point current fluctuations at $x=0$:
\begin{eqnarray}
\frac{1}{l} \left| \frac{\partial u}{\partial x} \right|_{x=0, \rm rms} & = & \sqrt{\frac{\hbar}{Xc\omega_n}} \frac{k_n \sin (k_n X)}{l}
\nonumber \\
& = & \sqrt{\frac{\hbar k_n^2}{X l^2 c \omega_n}} \times \sqrt{\frac{\tan^2(k_n X)}{1 + \tan^2(k_n X)}}
\nonumber \\
& = & \sqrt{\frac{\hbar\omega_n}{Xl}} \times \frac{1}{\sqrt{1 + \left( \frac{\omega_n}{\omega_{\rm cutoff}} \right)^2}}
\nonumber \\
& = & \frac{1}{Xl} \sqrt{\frac{\hbar \pi Z_0}{2}} \times \sqrt{\frac{\omega_n/\omega_0}{1 + \left( \frac{\omega_n}{\omega_{\rm cutoff}} \right)^2}}.
\label{Eq:CurrentZPF}
\end{eqnarray}
This expression can be substituted in the formula
\begin{equation}
\hbar g_n = L_c \times I_{\rm qubit} \times \frac{1}{l} \left| \frac{\partial u}{\partial x} \right|_{x=x_{\rm qubit}, \rm rms}
\label{Eq:CouplingStrengthFormula}
\end{equation}
to obtain the coupling strength $g_n$ as a function of mode frequency $\omega_n$. Here $g_n$ is the parameter that enters in the quantum Rabi model (QRM) Hamiltonian
\begin{equation}
\mathcal{H}_{QRM} = -\frac{\Delta_0}{2} \sigma_{x} - \frac{\epsilon}{2} \sigma_z + \sum_{n=0}^{\infty} \left\{ \hbar \omega_n \left( a_n^{\dagger} a_n + \frac{1}{2} \right) + \hbar g_n \sigma_z \left( a_n + a_n^{\dagger} \right) \right\},
\label{Eq:HamiltonianQRM}
\end{equation}
where $\Delta_0$ is the bare qubit gap, i.e.~not including renormalization by the Lamb shift, $\epsilon$ is the qubit bias, $\sigma_x$ and $\sigma_z$ are qubit Pauli operators, and $a_n$ and $a_n^{\dagger}$ are, respectively, the annihilation and creation operators for mode $n$. The coupling strength $g_n$ grows as $\sqrt{\omega_n}$ for $\omega\ll\omega_{\rm cutoff}$ and decreases as $1/\sqrt{\omega_n}$ for $\omega\gg\omega_{\rm cutoff}$, as found in several recent studies.

\subsection{Dependence of coupling strength on mode frequency and coupling inductance}

\begin{figure}[h]
\includegraphics[width=11.0cm]{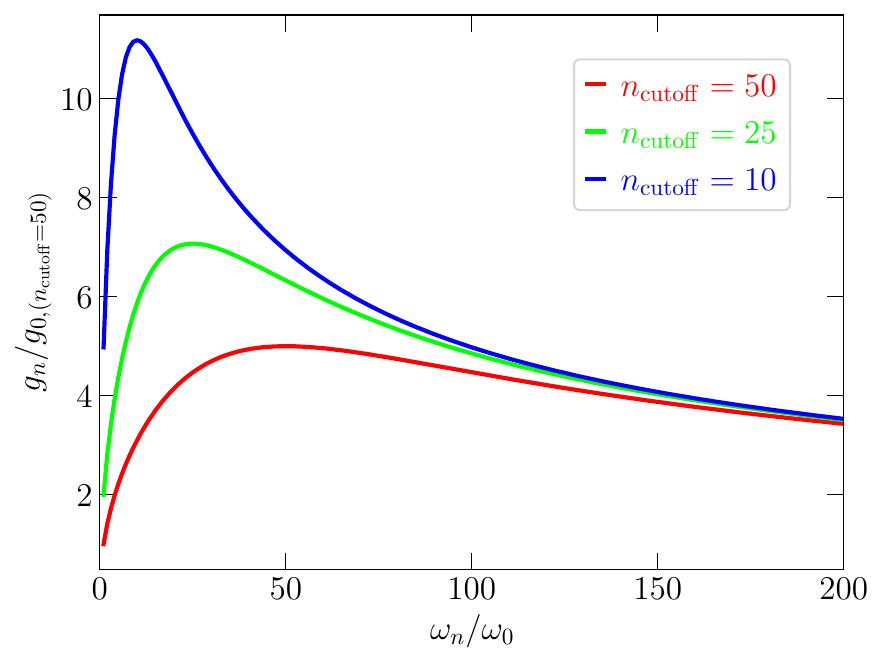}
\caption{Coupling strength $g_n$ as a function of mode frequency $\omega_n$ for three different values of the coupling inductance $L_c$. The different $L_c$ values are identified by the resulting $n_{\rm cutoff}$ values. Here it is assumed that all system parameters other than $L_c$ are kept fixed and $L_2 \gg L_c$, such that $n_{\rm cutoff} \propto 1/L_c$. The mode frequency $\omega_n$ is measured relative to the fundamental mode frequency $\omega_0$. The coupling strength $g_n$ is measured relative to $g_0$ for the $L_c$ value that gives $n_{\rm cutoff}=50$.}
\label{Fig:CouplingStrengthVsFrequency}
\end{figure}

We now consider the dependence of $g_n$ on $\omega_n$. Since the coupling between the qubit and the TLR arises from the shared coupling inductance $L_c$, it is interesting to also ask how $g_n$ depends on $L_c$. There is a factor $L_c$ that appears explicitly in Eq.~(\ref{Eq:CouplingStrengthFormula}). Both qubit and TLR currents also depend on $L_c$ in principle, because the same inductance that mediates the coupling can be seen as part of the qubit loop and part of the TLR circuit. For example, following Ref.~\cite{Yoshihara2022}, the qubit's persistent current is given by
\begin{equation}
I_{\rm qubit} \sim \frac{\Phi_0}{\pi (L_c+L_2)}.
\end{equation}
In other words, the flux qubit's persistent current $I_{\rm qubit}$ depends on $L_c$. However, if we consider the weak-coupling regime with $L_c \ll L_2$, and $L_c \ll Xl$, the qubit and low-frequency mode currents are independent of $L_c$ to lowest order. This approximation gives for the low-frequency modes
\begin{eqnarray}
\hbar g_n & \sim & L_c \times I_{\rm qubit} \times \frac{1}{Xl} \sqrt{\frac{\hbar \pi Z_0}{2}} \times \sqrt{\frac{\omega_n}{\omega_0}} \left( 1 - \frac{1}{2} \left( \frac{\omega_n}{\omega_{\rm cutoff}} \right)^2 \right)
\nonumber \\
& = & L_c \times I_{\rm qubit} \times \sqrt{\frac{\hbar\omega_n}{Xl}} \times \left( 1 - \frac{1}{2} \left( \frac{\omega_n}{\omega_{\rm cutoff}} \right)^2 \right).
\end{eqnarray}
In other words, $g_n\propto\sqrt{\omega_n}$, with relative corrections on the order of $(\omega_n/\omega_{\rm cutoff})^2$. This result is consistent with the low-frequency behavior obtained in previous studies. Additionally, we note that $g_n \propto L_c$. This result is intuitively to be expected, considering that $L_c$ is the mutual coupling inductance. This behaviour is illustrated in Fig.~\ref{Fig:CouplingStrengthVsFrequency}.

For the high-frequency modes
\begin{eqnarray}
\frac{1}{l} \left| \frac{\partial u}{\partial x} \right|_{x=0, \rm rms} & \sim & \frac{1}{Xl} \sqrt{\frac{\hbar \pi Z_0}{2}} \times \sqrt{\frac{\omega_{\rm cutoff}^2}{\omega_n \omega_0}}
\nonumber \\
& = & \frac{1}{XlL_c} \sqrt{\frac{\hbar \pi Z_0^3}{2 \omega_n \omega_0}}
\nonumber \\
& = & \frac{1}{L_c} \sqrt{\frac{\hbar Z_0^2}{Xl\omega_n}},
\end{eqnarray}
which gives
\begin{equation}
\hbar g_n \sim I_{\rm qubit} \sqrt{\frac{\hbar Z_0^2}{Xl\omega_n}} = I_{\rm qubit} \sqrt{\frac{2\hbar Z_0 \omega_0}{\pi\omega_n}}.
\label{Eq:CouplingStrengthAtHighFrequency}
\end{equation}
In agreement with past studies, $g_n\propto 1/\sqrt{\omega_n}$. Furthermore, Eq.~(\ref{Eq:CouplingStrengthAtHighFrequency}) reveals that the coupling strength is independent of the mutual inductance $L_c$ for the high-frequency modes, as can be seen in Fig.~\ref{Fig:CouplingStrengthVsFrequency}. This result is quite interesting. It means that we cannot increase the coupling strength to high frequency modes by increasing the qubit-TLR mutual inductance. Instead, treating $Z_0$ and $\omega_n$ as constants, the two possible approaches to increase $g_n$ are either increasing $I_{\rm qubit}$ or reducing the total TLR inductance $Xl$.

Note that in this section we assumed that $L_c$ is small compared to other inductances in the circuit in order to be able to derive simple expressions. Taking large $L_c$ would significantly modify the behaviour of the dynamical variables, which can lead to significant deviations from the approximate expressions that we derived. This situation could require numerical analysis of the equations of motion etc. We will not perform any such analysis in this paper.

\subsection{Modified frequency pattern for the low-frequency modes}

In the absence of the qubit, the mode frequencies of a quarter-wavelength TLR follow the pattern: $\omega_1/\omega_0=3$, $\omega_2/\omega_0=5$, ..., i.e.~$\omega_n/\omega_0=(2n+1)$. We now consider the correction to this pattern induced by the finiteness of $L_{c2}/Xl$. The motivation to look for such deviations is that a modified pattern could serve as an experimental test for the theory and a means to infer system parameters from measured spectra.

As can be seen from Eq.~(\ref{Eq:LowFrequencySolutionToFirstOrder}), the pattern $\omega_n/\omega_0=(2n+1)$ for the low-frequency modes is preserved at the lowest order considered in Sec.~\ref{Sec:Solutions}A. The corrections that we would like to calculate can come from one of two sources. If we start from Eq.~(\ref{Eq:NormalModeEquation_k}), including the next order in the expansion of the cotangent function in Eq.~(\ref{Eq:LowFrequencyNormalModeEquation}) will lead to frequency ratio corrections. Alternatively, starting from Eq.~(\ref{Eq:NormalModeEquationIncludingPhiTerm}) will give corrections even without going to higher orders in the cotangent function expansion. For the sake of clarity, we treat these two effects separately.

First, we take Eq.~(\ref{Eq:NormalModeEquation_k}) and use the expansion: $\cot(n\pi+\pi/2-\delta)\approx\delta + \delta^3/3$, i.e.
\begin{eqnarray}
n \pi + \frac{\pi}{2} - \widetilde{k_n X} & = & \frac{Xl}{L_{c2}} \cot (n\pi + \frac{\pi}{2} - \widetilde{k_n X})
\nonumber \\
& \approx & \frac{Xl}{L_{c2}} \left( \widetilde{k_n X} + \frac{(\widetilde{k_n X})^3}{3} \right).
\end{eqnarray}
This equation can be rearranged into the form
\begin{equation}
\widetilde{k_n X} \approx \frac{1}{1+Xl/L_{c2}} \left( n \pi + \frac{\pi}{2} - \frac{Xl}{L_{c2}} \frac{(\widetilde{k_n X})^3}{3} \right),
\label{Eq:knXIntermediateApproximateFormula}
\end{equation}
We now take the lowest-order approximation for $\widetilde{k_n X}$, i.e.~$\widetilde{k_n X}\approx (n\pi + \pi/2) / (1+Xl/L_{c2})$, and we substitute it in the cubic term on the right-hand side of Eq.~(\ref{Eq:knXIntermediateApproximateFormula}) to obtain
\begin{equation}
\widetilde{k_n X} \approx \frac{n \pi + \frac{\pi}{2}}{1+Xl/L_{c2}} - \frac{1}{3} \frac{Xl/L_{c2}}{(1+Xl/L_{c2})^4} \left( n \pi + \frac{\pi}{2} \right)^3,
\end{equation}
and therefore
\begin{equation}
k_n X \approx \left( n \pi + \frac{\pi}{2} \right) \left[ 1 - \frac{1}{1+Xl/L_{c2}} + \frac{1}{3} \frac{Xl/L_{c2}}{(1+Xl/L_{c2})^4} \left( n \pi + \frac{\pi}{2} \right)^2 \right].
\end{equation}
This expression then leads to the approximate relation
\begin{eqnarray}
\frac{\omega_n}{\omega_0} = \frac{k_n}{k_0} & \approx & (2n+1) \times \left[ 1 + \frac{1}{3} \frac{\left( n \pi + \frac{\pi}{2}\right)^2}{(Xl/L_{c2})^3} - \frac{1}{3} \frac{\left( \frac{\pi}{2}\right)^2}{(Xl/L_{c2})^3} \right]
\nonumber \\
& = & (2n+1) \times \left[ 1 + \frac{\pi^2}{3} \left( \frac{L_{c2}}{Xl} \right)^3 \left( n^2 + n \right) \right]
\nonumber \\
& = & (2n+1) \times \left[ 1 + \frac{8 \left( n^2 + n \right)}{3\pi n_{\rm cutoff}^3} \right].
\label{Eq:ModifiedFrequencyRatioCubic}
\end{eqnarray}
For the lowest three modes, we find that $\omega_1/\omega_0 \approx 3 + 16/(\pi n_{\rm cutoff}^3)$ and $\omega_2/\omega_0 \approx 5 + 80/(\pi n_{\rm cutoff}^3)$. In Ref.~\cite{Ao}, $n_{\rm cutoff}=13.2$, which gives $\omega_1/\omega_0 - 3 \approx 0.002$ and $\omega_2/\omega_0 - 5 \approx 0.01$. In other words, for $\omega_0/(2\pi)\approx 2.5$ GHz, we obtain $(\omega_1 - 3 \omega_0)/(2\pi) \approx 5$ MHz and $(\omega_2 - 5 \omega_0)/(2\pi) \approx 25$ MHz. Since the measurement accuracy for resonance frequencies in spectroscopy measurements is typically well below 1 MHz, a deviation of 5 MHz should be measurable experimentally.

Next, we consider Eq.~(\ref{Eq:NormalModeEquationIncludingPhiTerm}). If we take the limit $\omega_n\ll\omega_q$, Eq.~(\ref{Eq:LowFrequencyNormalModeEquation}) is replaced by
\begin{eqnarray}
n \pi + \frac{\pi}{2} - \widetilde{kX} & \approx & \frac{Xl}{L_{c2}} \widetilde{kX} \times \left( 1 + \frac{\mu}{\frac{(n \pi + \frac{\pi}{2} - \widetilde{kX})^2}{X^2cl\omega_q^2} - 1} \right)
\nonumber \\
& \approx & \frac{Xl}{L_{c2}} \widetilde{kX} \times \left( 1 - \mu  \left[ 1 + \frac{(2n + 1)^2 \omega_0^2}{\omega_q^2} \right] \right).
\label{Eq:LowFrequencyNormalModeEquationIncludingPhiTerm}
\end{eqnarray}
This equation can be rearranged into the form
\begin{equation}
\widetilde{k_n X} \approx \frac{n \pi + \frac{\pi}{2}}{1+\frac{Xl}{L_{c2}}\left(1 - \mu \left[ 1 + \frac{(2n + 1)^2 \omega_0^2}{\omega_q^2} \right] \right)},
\end{equation}
which then gives
\begin{equation}
k_n X \approx \left( n \pi + \frac{\pi}{2} \right) \left[ 1 - \frac{1}{1+\frac{Xl}{L_{c2}}(1-\mu)} - \frac{\mu(Xl/L_{c2})(2n + 1)^2 \omega_0^2}{\left(1+\frac{Xl}{L_{c2}}(1-\mu)\right)^2\omega_q^2} \right].
\end{equation}
We then obtain the approximate relation
\begin{eqnarray}
\frac{\omega_n}{\omega_0} = \frac{k_n}{k_0} & \approx & (2n+1) \times \left[ 1 - \frac{2}{\pi n_{\rm cutoff}(1-\mu)^2} \left( \frac{\mu(2n + 1)^2 \omega_0^2}{\omega_q^2} - \frac{\mu \omega_0^2}{\omega_q^2} \right) \right]
\nonumber \\
& \approx & (2n+1) \times \left[ 1 - \frac{8\mu \omega_0^2}{\pi n_{\rm cutoff} (1-\mu)^2 \omega_q^2} (n^2 + n) \right].
\end{eqnarray}
Taking $n_{\rm cutoff}=13.2$, $\mu=0.1$ and $\omega_q/\omega_0 = 20$, we find that $\omega_1/\omega_0 - 3 \approx -4 \times 10^{-4}$ and $\omega_2/\omega_0 - 5 \approx -2 \times 10^{-3}$. These values are smaller than the ones calculated based on Eq.~(\ref{Eq:ModifiedFrequencyRatioCubic}) by about one order of magnitude. They should therefore not significantly affect the frequency relations. One potentially important effect that appears when we use Eq.~(\ref{Eq:NormalModeEquationIncludingPhiTerm}) is the fact that the factor $Xl/L_{c2}$ in Eq.~(\ref{Eq:LowFrequencyNormalModeEquationIncludingPhiTerm}) is now multiplied by the factor $1-\mu$ for the low-frequency modes. Since Eq.~(\ref{Eq:ModifiedFrequencyRatioCubic}) has the factor $(Xl/L_{c2})^{-3}$, the frequency deviations calculated from Eq.~(\ref{Eq:ModifiedFrequencyRatioCubic}) will be amplified by the factor $(1-\mu)^{-3}$, which is approximately equal to 1.3 for the parameters that we are using in our estimates.

In the context to calculating small variations in $\omega_n/\omega_0$, we consider the effect of having a finite $C_R$ on the mode frequencies. For this calculation, we ignore the presence of the qubit and focus on the effect of the finite $C_R$. The wave equation for $\phi$ is still given by Eq.~(\ref{Eq:phiPreRedefinition}), and its solution can still be expressed as
\begin{equation}
\phi(x,t) = u_c e^{-i\omega t} \cos (k [x-x_0]),
\end{equation}
with $\omega=k/\sqrt{cl}$. In the absence of the qubit, the boundary condition at $x=0$ is $\phi(x)|_{x=0}=0$, which allows us to express $\phi(x,t)$ as
\begin{equation}
\phi(x,t) = u_c e^{-i\omega t} \sin (k x).
\label{Eq:BareTLRWaveEquationSolution}
\end{equation}
The last line of Eq.~(\ref{Eq:2ndOEOMDiscretephi}), along with the $e^{-i\omega t}$ time dependence of $\phi(x,t)$, now gives the boundary condition:
\begin{equation}
\left( -\omega^2 \phi + \frac{1}{lC_R} \frac{d\phi}{dx} \right) \Bigg|_{x=X} = 0.
\label{Eq:BoundaryConditionCR}
\end{equation}
Substituting Eq.~(\ref{Eq:BareTLRWaveEquationSolution}) into Eq.~(\ref{Eq:BoundaryConditionCR}) gives the transcendental equation
\begin{equation}
k \tan (k X) = \frac{c}{C_R}.
\end{equation}
This equation has exactly the same form as Eq.~(\ref{Eq:NormalModeEquation_k}), but with $c/C_R$ instead of $l/L_{c2}$ on the right-hand side of the equation. Considering the case $C_R/(Xc) \ll 1$, we can use the same derivation as the one used earlier in this subsection and obtain
\begin{equation}
\frac{\omega_n}{\omega_0} \approx (2n+1) \times \left[ 1 + \frac{\pi^2}{3} \left( \frac{C_R}{Xc} \right)^3 \left( n^2 + n \right) \right].
\end{equation}
The effect of a finite $C_R$ is therefore to modify the ratios $\omega_n/\omega_0$ in the same way that the coupling to the qubit modifies the frequency ratios. Depending on the relationship between the two small parameters $L_{c2}/(Xl)$ and $C_R/(Xc)$, either one of the two mechanisms can be dominant.

The design parameters of Ref.~\cite{Ao} are $X=10.75$ mm, $l=437$ nH/m, $c=162$ pF/m, $L_c=231$ pH, $L_2 = 823$ pH and $C_R = 3.46 \times 10^{-4}$ pF. These parameters give $L_{c2}/(Xl)\sim 4 \times 10^{-2}$ and $C_R/(Xc)\sim 2 \times 10^{-4}$, which suggests that the effect of the qubit will be much larger than the effect of the finite $C_R$. It should be noted, however, that the coupling of the TLR to the measurement transmission line in Ref.~\cite{Ao} was via an additional inductance that is not included in our theoretical model. As a result, the capacitance $C_R$ could in principle be made very small in that experiment without affecting the coupling between the TLR and the probe signal. If the circuit is designed such that the TLR is probed via the capacitor $C_R$, one might need to increase this capacitance and take more care in treating the effects of the finite $C_R$.

Here it is worth establishing the relation between the ratio $C_R/(Xc)$ and the TLR's quality factor $Q_{\rm TLR}$ (not to be confused with the charge variable $Q$ introduced in Sec.~\ref{Sec:Setup}), because $Q_{\rm TLR}$ can be measured more directly than other circuit parameters. Following Ref.~\cite{Goeppl}, and assuming that all dissipation in the TLR is via the $C_R$ capacitor, $Q_{\rm TLR}$ is given by
\begin{equation}
Q_{\rm TLR} = \frac{X c}{4 \omega_0 Z_0 C_R^2} = \frac{1}{4 \omega_0 Z_0 X c} \left( \frac{X c}{C_R} \right)^2.
\end{equation}
Using the last formula for $\omega_0$ in Eq.~(\ref{Eq:FundamentalModeFrequency}), we find that $4 \omega_0 Z_0 X c = 2 \pi$. As a result, we obtain the formula
\begin{equation}
Q_{\rm TLR} = \frac{1}{2\pi} \left( \frac{Xc}{C_R} \right)^2,
\end{equation}
or, in other words,
\begin{equation}
\frac{C_R}{Xc} = \frac{1}{\sqrt{2 \pi Q_{\rm TLR}}}.
\end{equation}
If, for example, we take the relatively low quality factor $Q_{\rm TLR}=10^3$, we obtain the relatively high estimate $C_R/(Xc)=0.013$, which is still significantly lower than the value $L_{c2}/(Xl)=0.048$ for the circuit of Ref.~\cite{Ao}. Since these small factors are raised to the third power in the formula for $\omega_n/\omega_0$, the effect of the qubit should be more than an order of magnitude larger than the effect of the finite $C_R$ in a realistic setup.

\section{Lamb shift}
\label{Sec:LambShift}

One of the important questions in the study of multimode cavity QED is the Lamb shift and its convergence as we take into account the coupling between the qubit and an increasingly large number of modes.

If the qubit frequency is small relative to the oscillator frequency, the renormalized gap, i.e.~the Lamb shifted qubit frequency, is given by
\begin{equation}
\Delta = \Delta_0 \exp \left\{ - 2 \sum_n \left( \frac{g_n}{\omega_n} \right)^2 \right\},
\end{equation}
which is the straightforward generalization of the single-mode formula \cite{Ashhab2010}
\begin{equation}
\Delta = \Delta_0 \exp \left\{ - 2 \left( \frac{g}{\omega} \right)^2 \right\}.
\end{equation}
It should be emphasized that the above formula is valid in a wide range of parameters. For example, if we think of the normalization process as occurring in steps starting at $n\to\infty$ and gradually going down to $n=0$, the above formula is valid as long as the renormalized value of $\Delta$ is smaller than $\omega_n$ in every step of the process. Using Eq.~(\ref{Eq:CurrentZPF}) to infer the functional dependence of $g_n$ on $\omega_n$, we can express the formula for $\Delta$ as
\begin{equation}
\Delta \approx \Delta_0 \exp \left\{ - 2 \left( \frac{g_0}{\omega_0} \right)^2 \sum_{n=1,3,5,...} \frac{1}{n\left( 1 + \frac{n^2}{n_{\rm cutoff}^2}\right)} \right\}.
\label{Eq:LambShiftOddN}
\end{equation}

Using the software package Mathematica, we found that the sum in Eq. (\ref{Eq:LambShiftOddN}) is given by
\begin{equation}
\sum_{n=1,3,5,...} \frac{1}{n\left( 1 + \frac{n^2}{n_{\rm cutoff}^2}\right)} = \frac{\gamma + 2\log 2}{2} + \frac{\psi (\frac{1+i n_{\rm cutoff}}{2}) + \psi (\frac{1-i n_{\rm cutoff}}{2})}{4},
\end{equation}
where $\gamma$ is Euler's constant (approximately 0.577), and $\psi (x)$ is the digamma function. For large $n_{\rm cutoff}$, the above expression reduces to the simpler function:
\begin{eqnarray}
\lim_{n_{\rm cutoff}\to\infty} \left[ \sum_{n=1,3,5,...} \frac{1}{n\left( 1 + \frac{n^2}{n_{\rm cutoff}^2}\right)} \right] & = & \frac{\gamma + \log 2}{2} + \frac{\log n_{\rm cutoff}}{2}
\nonumber \\
& \approx & 0.635 + 0.5 \log n_{\rm cutoff}.
\label{Eq:LambShiftSumOdd}
\end{eqnarray}
In other words, the expression $0.635 + 0.5 \log n_{\rm cutoff}$ provides a good approximation for the sum, as long as $n_{\rm{cutoff}} \gg 1$.

Here it is worth making a small digression and giving the related sum
\begin{equation}
\sum_{n=1,2,3,...} \frac{1}{n\left( 1 + \frac{n^2}{n_{\rm cutoff}^2}\right)} = \gamma + \frac{\psi (1+i n_{\rm cutoff}) + \psi (1-i n_{\rm cutoff})}{2}.
\end{equation}
For large $n_{\rm cutoff}$, we obtain the asymptotic behavior
\begin{equation}
\lim_{n_{\rm cutoff}\to\infty} \left[ \sum_{n=1,2,3,...} \frac{1}{n\left( 1 + \frac{n^2}{n_{\rm cutoff}^2}\right)} \right] = \gamma + \log n_{\rm cutoff}.
\label{Eq:LambShiftSumOddEven}
\end{equation}
This formula can be relevant to a situation in which a qubit is capacitively coupled to a half-wavelength TLR with capacitors at both ends, and the qubit is placed at one of the TLR's ends to maximize the coupling to all modes. The differences between Eqs.~(\ref{Eq:LambShiftSumOdd}) and (\ref{Eq:LambShiftSumOddEven}) are intuitively logical: a slightly different constant term and a factor of 2 in the $\log n_{\rm cutoff}$ term.

The slow logarithmic dependence of the above sums on $n_{\rm cutoff}$ means that $\Delta$ changes very slowly as a function of $n_{\rm cutoff}$. It also means that the sum remains on the order of 1 and can be treated as almost constant compared to other factors that play a role in determining the Lamb shift. For example, for $n_{\rm cutoff}=100$, the sum is approximately equal to 3.

\section{Conclusion}
\label{Sec:Conclusion}

We have performed theoretical analysis of a circuit comprising a flux qubit inductively coupled to a quarter-wavelength TLR. We showed how the qubit naturally decouples from the high-frequency modes of the TLR. Our results on this point agree with past results on somewhat similar circuits. Our analysis therefore complements previous theoretical studies and adds insight into the physics of the decoupling effect. We also derived new formulae for the mode frequencies, coupling strengths and Lamb shift. By avoiding certain approximations used in previous studies, we derived more accurate results, which allow us to predict previously unknown features in the spectrum of the system. Our results can help guide future experiments to test the physics of ultrastrong and deep-strong coupling in multimode circuit QED.

\section*{Appendix A: Deriving the equations of motion for the TLR variables}

In this Appendix, we provide a more detailed calculation for how we can decouple the equations of motion for the variables $\Phi$ and $\phi$, i.e.~how we can derive Eqs.~(\ref{Eq:phiPostRedefinition}-\ref{Eq:BoundaryConditionXPostRedefinition}) and Eq.~(\ref{Eq:NormalModeEquationIncludingPhiTerm}) from Eqs.~(\ref{Eq:phiPreRedefinition}-\ref{Eq:BoundaryConditionXPreRedefinition}).

If we want to find the oscillation modes of a system with continuous variables in some multi-dimensional trapping potential, we first find the ground state. We can find the semiclassical ground state from Eqs.~(\ref{Eq:PhiPreRedefinition},\ref{Eq:phiPreRedefinition}) by requiring that all the dynamical variables be constant in time, i.e.~looking for stationary solutions for the equations of motion. Setting $\dot{\Phi}=\partial \phi/\partial t = 0$ in Eqs.~(\ref{Eq:PhiPreRedefinition},\ref{Eq:phiPreRedefinition}), we obtain the stationary-state equations
\begin{eqnarray}
- \frac{1}{C_q} \frac{dU_q(\Phi,\Phi_{\rm ext})}{d\Phi} - \frac{\Phi - \phi(x=0)}{C_q L_2} & = & 0 \\
\frac{1}{cl} \frac{\partial^2 \phi}{\partial x^2} & = & 0 \\
\left( \frac{1}{cl} \frac{\partial \phi}{\partial x} - \frac{\phi}{c L_c} + \frac{\Phi - \phi}{c L_2} \right) \Bigg|_{x=0} & = & 0 \\
\frac{\partial \phi}{\partial x} \Bigg|_{x=X} & = & 0,
\end{eqnarray}
which in turn give
\begin{eqnarray}
\phi(x) & = & \phi_{GS} {\rm \ (i.e. \ constant \ independent \ of} \ x) \\
\frac{dU_q(\Phi,\Phi_{\rm ext})}{d\Phi} \Bigg|_{\Phi_{GS}} + \frac{\Phi_{GS}}{L_2} & = & \frac{\phi_{GS}}{L_2}
\label{Eq:PhiGS} \\
\phi_{GS} & = & \frac{L_{c2} \Phi_{GS}}{L_2}.
\label{Eq:phiGS}
\end{eqnarray}
Substituting Eq.~(\ref{Eq:phiGS}) in Eq.~(\ref{Eq:PhiGS}), we obtain
\begin{equation}
\frac{dU_q(\Phi,\Phi_{\rm ext})}{d\Phi} \Bigg|_{\Phi_{GS}} + \left( \frac{1}{L_2} - \frac{L_{c2}}{L_2^2} \right) \Phi_{GS} = 0.
\label{Eq:PhiGSFinal}
\end{equation}

In this work, we are assuming that the variable $\Phi$ describes a flux qubit. Then there will in general be two solutions, i.e.~two values of $\Phi_{GS}$ that satisfy Eq.~(\ref{Eq:PhiGSFinal}), determined to a large extent by the function $U_q(\Phi,\Phi_{\rm ext})$. Focusing on one of these two solutions (and assuming that $\Phi$ is nonzero, e.g.~$\Phi_{GS}\sim\Phi_0/4$), the equation for $\phi_s$ gives a nonzero constant for $\phi(x)$. We can now look for deviations of $\Phi$ and $\phi$ away from the ground state values [and we call the deviations $\delta\Phi(t)$ and $\delta\phi(x,t)$] and analyze their dynamics. Then we find the equations of motion
\begin{eqnarray}
\ddot{\delta\Phi} & = & - \left( \frac{1}{C_q} \frac{d^2U_q(\Phi,\Phi_{\rm ext})}{d\Phi^2} \Bigg|_{\Phi_{GS}} + \frac{1}{C_q L_2} \right) \delta\Phi + \frac{\delta\phi(x=0,t)}{C_q L_2}
\label{Eq:2ndOEOMdeltaPhi} \\
\frac{\partial^2 \delta\phi}{\partial t^2} & = & \frac{1}{cl} \frac{\partial^2 \delta\phi}{\partial x^2},
\end{eqnarray}
with boundary conditions
\begin{eqnarray}
\left( \frac{1}{cl} \frac{\partial \delta\phi}{\partial x} - \frac{\delta\phi}{c L_c} + \frac{\delta\Phi - \delta\phi}{c L_2} \right) \Bigg|_{x=0} & = & 0
\label{Eq:BoundaryCondition0deltaphi} \\
\frac{\partial \delta\phi}{\partial x} \Bigg|_{x=X} & = & 0.
\label{Eq:BoundaryConditionXdeltaphi}
\end{eqnarray}

The equations of motion are now both linear differential equations, as is expected in calculations of normal modes. If we ignore $\delta\Phi$ in Eq.~(\ref{Eq:BoundaryCondition0deltaphi}), we obtain Eqs.~(\ref{Eq:phiPostRedefinition}-\ref{Eq:BoundaryConditionXPostRedefinition}). These equations do in fact provide a generally good approximation for the TLR's normal modes while remaining simple. We therefore use these equations as the starting point in several derivations in the main text. We can, however, proceed without ignoring $\delta\Phi$ in Eq.~(\ref{Eq:BoundaryCondition0deltaphi}), as we show next.

First, we rewrite Eq.(\ref{Eq:2ndOEOMdeltaPhi}) as:
\begin{equation}
\ddot{\delta\Phi} + \omega_q^2 \delta\Phi = \frac{\delta\phi(x=0,t)}{C_q L_2},
\label{Eq:ForcedHO}
\end{equation}
where $\omega_q=1/\sqrt{C_q L_{q2}}$, $L_{q2}=L_q L_2 / (L_q + L_2)$, and $L_q=\left(d^2U_q(\Phi,\Phi_{\rm ext})/d\Phi^2\right) |_{\Phi_{GS}}$. It should be noted that $\omega_q$ is not the frequency separation between the qubit's 0 and 1 states, but rather the classical oscillation frequency about the appropriate local minimum in the effective potential for the variable $\Phi$. In particular, the flux qubit's minimum gap is typically a few GHz and is much smaller than $\omega_q$, which is typically a few tens of GHz. Equation (\ref{Eq:ForcedHO}) is that of a forced harmonic oscillator, with the term on the right-hand side playing the role of the driving force. If we focus on the TLR mode with frequency $\omega_n$, such that $\delta\phi(x=0,t)=A e^{-i\omega_n t}$, the solution of Eq.~(\ref{Eq:ForcedHO}) is given by $\delta\Phi (t) = B e^{-i\omega_n t}$, with
\begin{equation}
B = \frac{A}{C_q L_2 \left( \omega_q^2 - \omega_n^2 \right)}.
\label{Eq:ForcedHOSolutionDeltaPhi}
\end{equation}
Substituting Eq.~(\ref{Eq:ForcedHOSolutionDeltaPhi}) into Eq.~(\ref{Eq:BoundaryCondition0deltaphi}), we obtain the modified boundary condition
\begin{equation}
\left( \frac{1}{cl} \frac{\partial \delta\phi}{\partial x} - \frac{\delta\phi}{c L_{c2}} + \frac{1}{C_q L_2 \left( \omega_q^2 - \omega_n^2 \right)} \frac{\delta\phi}{c L_2} \right) \Bigg|_{x=0} = 0.
\label{Eq:BoundaryCondition0deltaphiIncludingPhiTerm}
\end{equation}
Equation (\ref{Eq:BoundaryCondition0deltaphiIncludingPhiTerm}) differs from Eq.~(\ref{Eq:BoundaryCondition0PostRedefinition}) in the presence or absence of the last term on the left-hand side. If we include this term, Eq.~(\ref{Eq:NormalModeEquation_k}) is replaced by Eq.~(\ref{Eq:NormalModeEquationIncludingPhiTerm}).

A few comments are in order regarding Eq.~(\ref{Eq:NormalModeEquationIncludingPhiTerm}). We generally assume that $L_c \ll L_2$ and $L_q \lesssim L_2$. As a result, $\mu\ll 1$. The factor $k^2/(cl\omega_q^2)-1$ can be expressed as $\omega^2/\omega_q^2-1$, which indicates that there will be a resonance effect. The TLR modes that are close in frequency to $\omega_q$ will be the ones that are most strongly modified by the additional term in Eq.~(\ref{Eq:NormalModeEquationIncludingPhiTerm}) when compared to Eq.~(\ref{Eq:NormalModeEquation_k}). The mode frequencies slightly below $\omega_q$ are pushed down, while the frequencies slightly above $\omega_q$ are pushed up, as is expected for coupled harmonic oscillators. Far away from the resonance point, the last term in Eq.~(\ref{Eq:NormalModeEquationIncludingPhiTerm}) can be ignored, and Eq.~(\ref{Eq:NormalModeEquation_k}) provides a good approximation for the TLR mode frequencies. In fact, even though the blue and green lines in Fig.~\ref{Fig:ModeFrequencyTranscendentalEquation} seem to be significantly different, the mode frequencies (given by the x-axis values obtained from the points of intersection with the red lines) are not significantly different between the two cases. Another point worth noting here is that the $1/x$-like feature in the green line results in one additional solution to Eq.~(\ref{Eq:NormalModeEquationIncludingPhiTerm}) compared to Eq.~(\ref{Eq:NormalModeEquation_k}). This additional mode is associated with qubit oscillations. It always occurs close to $\omega_q$, as expected.

\section*{Appendix B: Coupling strength between Qubit and TLR modes}

In Sec.~\ref{Sec:Solutions}.D, we derived the coupling strength using the standard formula for magnetic coupling between two current-carrying wires, i.e.~the product of the mutual inductance and the two relevant currents. In this Appendix, we follow an alternative derivation based on the circuit-variable description of the circuit, i.e.~using the variables $Q$, $q_j$, $\Phi$ and $\phi_n$. The motivation to perform this calculation is to follow the same approach used for most other calculations in this work and avoid hand-waving arguments as much as possible, especially considering that we are dealing with an inherently nonlinear system.

In the QRM, the flux qubit is approximated as a two-level system, with two quantum states that differ by the current going around the qubit loop. The coupling strength between the qubit and TLR modes can be calculated from the energies of the corresponding states of the electric circuit. These energies can be calculated from the TLR mode frequencies for the two qubit states, as well as the overlap between the mode wave functions for the two qubit states. For this purpose, we consider two flux qubit states with different values of $\Phi_{GS}$ as discussed in Appendix A. We consider the case where the mode frequencies are almost the same for both qubit states. Indeed, this condition is needed for the standard QRM Hamiltonian to be valid. Since we generally expect $L_q$ to be different for the two qubit states, we find that we must have $\mu\ll 1$ for this condition to be satisfied. Next we consider the overlap between the TLR mode ground state wave functions for the two different qubit states. The two qubit states with opposite persistent-current directions have two different values of $\Phi_{GS}$, which lead to two different values of $\phi_{GS}$, as described by Eq.~(\ref{Eq:phiGS}). In other words, the TLR ground-state wave functions for the two qubit states will be shifted relative to each other.

If we consider the QRM Hamiltonian in Eq.~(\ref{Eq:HamiltonianQRM}), as explained in Ref.~\cite{Ashhab2010}, the two $\sigma_z$ states of the qubit result in effective harmonic oscillator Hamiltonians with ground state field values that are separated by $2\alpha$, where $\alpha = g/\omega$. This separation should be measured relative to the spread of the ground-state wave function, i.e.~the vacuum fluctuations of the field variable. When the Hamiltonian is written in terms of the creation and annihilation operators, the formula is simple:
\begin{equation}
\left( a + a^{\dagger} \right)_{\rm rms} = \left\langle \left( a + a^{\dagger} \right)^2 \right\rangle^{1/2} = 1.
\end{equation}

To calculate the vacuum fluctuations of the TLR mode amplitudes, we write the Hamiltonian in Eq.~(\ref{Eq:HamiltonianCV}) in terms of the normal mode variables. The eigenvalue problem for the TLR modes is a Sturm-Liouville problem, which means that the eigenfunctions form a complete orthogonal set. Considering the mode functions derived in Sec.~\ref{Sec:Solutions}, we define the mode amplitudes $\varphi_n$ via the equation
\begin{equation}
\delta\phi(x) = \sum_{n=0}^{\infty} \cos \left( k_n [x-X] \right) \varphi_n,
\end{equation}
with the inverse
\begin{equation}
\varphi_n = \frac{1}{\int_{0}^{X} \cos^2 \left( k_n [x-X] \right) dx} \int_{0}^{X} \cos \left( k_n [x-X] \right) \delta\phi(x) dx.
\end{equation}
In particular, if we take a constant function $\delta\phi(x) = \overline{\varphi}$, we find that its amplitude components $\varphi_n$ in the different modes are given by
\begin{eqnarray}
\frac{\varphi_n}{\overline{\varphi}} & = & \frac{1}{\int_{0}^{X} \cos^2 \left( k_n [x-X] \right) dx} \int_{0}^{X} \cos \left( k_n [x-X] \right) dx
\nonumber \\
& = & \frac{1}{\frac{X}{2} \left( 1 + \frac{\sin (2k_nX)}{2k_nX} \right)} \frac{\sin(k_n X)}{k_n}
\nonumber \\
& \approx & \frac{2\sin(k_n X)}{k_n X}
\nonumber \\
& = & \frac{2}{k_n X} \sqrt{\frac{\tan^2 (k_n X)}{1 + \tan^2 (k_n X)}}
\nonumber \\
& = & \frac{2}{k_n X} \sqrt{\frac{1}{1 + \left( \frac{\omega_n}{\omega_{\rm cutoff}} \right)^2}}.
\end{eqnarray}
Considering Eq.~(\ref{Eq:phiGS}), we find that the two qubit states result in $\varphi_n$ values that are separated by a distance
\begin{equation}
\frac{L_{c2} \left(\Phi_{GS1}-\Phi_{GS2}\right)}{L_2} \times \frac{2}{k_n X} \sqrt{\frac{1}{1 + \left( \frac{\omega_n}{\omega_{\rm cutoff}} \right)^2}}.
\end{equation}

The relation between the momenta in the original Hamiltonian and the TLR mode momenta can be derived from the relation for the phase variables:
\begin{equation}
q_n = \frac{\partial \mathcal{L}}{\partial \dot{\varphi}_n} = \sum_{j=0}^{N} \cos \left( k_n \frac{j-N}{N} X \right) \frac{\partial \mathcal{L}}{\partial \dot{\phi}_j} = \sum_{j=0}^{N} \cos \left( k_n \frac{j-N}{N} X \right) \delta q_j,
\end{equation}
which implies that
\begin{equation}
\delta q_j = \delta x \sum_{n=0}^{N} \frac{1}{\int_{0}^{X} \cos^2 \left( k_n [x-X] \right) dx} \cos \left( k_n \frac{j-N}{N} X \right) q_n.
\end{equation}
Here we are using the upper limit $N$ to avoid counting more modes (with momenta $q_n$) than the original momentum variables $\delta q_j$, with the understanding that we will take the limit $N\to\infty$ in the next step. We also do not take the continuous limit for $\delta q_j$ to avoid dealing with intermediate-step infinities and their subsequent cancellation.

Focusing on the TLR terms in the Hamiltonian, we can write these as
\begin{eqnarray}
\sum_{j=0}^{N-1} \left( \frac{1}{2c \delta x} \delta q_j^2 + \frac{1}{2l \delta x} \left( \phi_j - \phi_{j+1} \right)^2 \right) & &
\nonumber \\
& & \hspace{-4cm} \to \sum_{n=0}^{\infty} \left( \frac{1}{2c \int_{0}^{X}\cos^2(k_n[x-X])dx} q_n^2 + \frac{k_n^2}{2l} \int_{0}^{X}\sin^2(k_n[x-X])dx \varphi_{n}^2 \right)
\nonumber \\ 
& & \hspace{-4cm} = \sum_{n=0}^{\infty} \left( \frac{1}{cX \left( 1 + \frac{\sin (2k_nX)}{2k_nX} \right)} q_n^2 + \frac{k_n^2 X}{4l} \left( 1 - \frac{\sin (2k_nX)}{2k_nX} \right) \varphi_{n}^2 \right).
\end{eqnarray}
We do not show the derivation in detail here, but the other terms in the Hamiltonian, which contain the variables $\varphi_n$ but not $q_n$, modify the above term in the Hamiltonian to
\begin{equation}
\sum_{n=0}^{\infty} \left( \frac{1}{cX \left( 1 + \frac{\sin (2k_nX)}{2k_nX} \right)} q_n^2 + \frac{k_n^2 X}{4l} \left( 1 + \frac{\sin (2k_nX)}{2k_nX} \right) \varphi_{n}^2 \right).
\end{equation}
As a consistency check, we note that this form of the Hamiltonian gives
\begin{equation}
\omega_n^2 = 4 \times \frac{1}{cX \left( 1 + \frac{\sin (2k_nX)}{2k_nX} \right)} \times \frac{k_n^2 X}{4l} \left( 1 + \frac{\sin (2k_nX)}{2k_nX} \right) = \frac{k_n^2}{cl},
\end{equation}
as expected. The term $\sin (2k_nX)/(2k_nX)$ is small, decays quickly with increasing $n$ and can therefore be ignored. As a result, the vacuum fluctuations of $\varphi_n$ can be approximated by
\begin{equation}
\varphi_{n,\rm rms} = \sqrt{\frac{\hbar}{X c \omega_n}} = \sqrt{\frac{2\hbar\omega_0 Z_0}{\pi\omega_n}}.
\end{equation}

Having obtained (1) the $n$th mode component of the separation between the two $\phi_{GS}$ values and (2) the vacuum fluctuations in $\varphi_n$, we calculate their ratio $\delta\alpha_n$:
\begin{eqnarray}
\delta\alpha_n & = & \frac{L_{c2} \left(\Phi_{GS1}-\Phi_{GS2}\right)}{L_2} \times \frac{2}{k_n X} \sqrt{\frac{1}{1 + \left( \frac{\omega_n}{\omega_{\rm cutoff}} \right)^2}} \times \sqrt{\frac{\pi\omega_n}{2\hbar\omega_0 Z_0}}
\nonumber \\
& = & L_{c} \times \frac{\left(\Phi_{GS1}-\Phi_{GS2}\right)}{L_c + L_2} \times \frac{2}{\omega_n X l} \times \sqrt{\frac{\pi\omega_n Z_0}{2\hbar\omega_0}} \times \sqrt{\frac{1}{1 + \left( \frac{\omega_n}{\omega_{\rm cutoff}} \right)^2}},
\end{eqnarray}
which can then be substituted in the formula $g_n=\delta\alpha_n \omega_n/2$ to obtain the coupling strength:
\begin{equation}
\hbar g_n = L_{c} \times \frac{\left(\Phi_{GS1}-\Phi_{GS2}\right)}{L_c + L_2} \times \frac{1}{Xl} \sqrt{\frac{\hbar\pi Z_0}{2}} \times \sqrt{\frac{\omega_n / \omega_0}{1 + \left( \frac{\omega_n}{\omega_{\rm cutoff}} \right)^2}}.
\end{equation}
We therefore recover the standard, intuitive formula for the coupling strength used in Sec.~\ref{Sec:Solutions}.D, i.e.~Eq.~(\ref{Eq:CouplingStrengthFormula}). This derivation also illustrates how the inductive energy is proportional to the mutual inductance $L_c$, in spite of the appearance of the $\Phi\phi_0$ cross term in Eq.~(\ref{Eq:HamiltonianCV}).

\section*{Acknowledgment}

This work was supported by Japan's Ministry of Education, Culture, Sports, Science and Technology (MEXT) Quantum Leap Flagship Program Grant Number JPMXS0120319794 and by Japan Science and Technology Agency Core Research for Evolutionary Science and Technology Grant Number JPMJCR1775. AL was supported by a NSERC Discovery grant.

\end{document}